\documentclass{ws-procs961x669}            
\def \gh{\vphantom{$\fbox{\Big[}$}}

\begin{document}
\begin{flushright}
DESY 16-134\\
July 2016
\end{flushright}
\title{Rare B-Meson Decays at the Crossroads \footnote{
To be published in the Proceedings of the Conference on New Physics at the Large Hadron Collider,
Nanyang Technological University, Singapore.  29 February - 4 March, 2016.}}

\author{Ahmed Ali$^*$}

\address{Deutsches Elektronen-Synchritron DESY\\
D-22607 Hamburg, Germany\\
$^*$E-mail: ahmed.ali@desy.de}

\begin{abstract}
 Experimental era of rare $B$-decays started with the  measurement of  $B \to K^* \gamma$
 by CLEO  in 1993, followed two years later by the measurement of the inclusive decay
 $B \to X_s \gamma$, which  serves as the standard candle in this field.
 The frontier  has moved in the meanwhile to the experiments at the LHC, in particular, LHCb, 
 with the decay $B^0 \to \mu^+ \mu^-$ at about 1 part in $10^{10}$  being the smallest branching fraction 
 measured so far. Experimental precision achieved in this area has put the standard model to unprecedented stringent tests and more are in the offing in the near future. I review some key measurements in radiative, semileptonic and leptonic rare $B$-decays,  contrast them with their estimates in the SM, and  focus on several mismatches reported recently.  They  are too numerous to be ignored, yet , standing alone, none of them is significant enough to warrant the breakdown of the  SM.  Rare $B$-decays find themselves at the crossroads, possibly pointing to new horizons, but quite likely requiring an improved theoretical description in the context of the SM. An independent 
 precision  experiment such as Belle II may help greatly in clearing some of the current experimental issues. 
\end{abstract}


\bodymatter

\section{Introduction}
The interest in studying  rare $B$  decays is immense. This is due to the circumstance that
these decays, such as $b \to (s,d) \;\gamma, \; b \to (s,d ) \; \ell^+\ell^-$, are flavour-changing-neutral-current (FCNC) processes, involving the quantum number  transitions $ | \Delta B |=1, |\Delta Q |=0$. In the SM \cite{Glashow:1961tr}, they are
not allowed at the tree level, but are  induced by  loops   and are  
 governed by the GIM (Glashow-Iliopoulos-Maiani) mechanism \cite{Glashow:1970gm}, which
imparts them sensitivity to higher masses,  $(m_t, m_W)$.
 As a consequence, they determine the  CKM  \cite{Cabibbo:1963yz} (Cabibbo-Kobayashi-Maskawa)
 matrix elements. Of these, the elements in the third row,
 $V_{td}$, $V_{ts}$ and $V_{tb}$ are of particular interest. While $\vert V_{tb}\vert$ has been measured in the
 production and decays of the top quarks in hadronic collisions \cite{Agashe:2014kda}, the first two are currently not yet directly
 accessible. In the SM, these CKM matrix elements have been indirectly determined from the $B^0$ - $\bar{B}^0$
 and $B_s^0$ - $\bar{B}_s^0$ mixings. Rare $B$-decays provide  independent measurements of the same
 quantities. 
 
  In theories involving physics beyond the SM (BSM), such as the 2-Higgs doublet models
 or supersymmetry, transitions involving the FCNC processes are 
 sensitive to the masses and couplings of the new particles. Precise experiments and
theory  are needed to establish or definitively rule out the BSM effects.
Powerful calculation techniques, such as the heavy quark effective theory (HQET) \cite{Manohar:2000dt}
and the soft collinear effective theory (SCET)
 \cite{Bauer:2000yr,Beneke:2002ph,Becher:2014oda}
 have been developed to incorporate power $1/m_c$ and $1/m_b$ corrections to the
perturbative QCD estimates. More importantly, they enable a better theoretical description by separating the
various scales involved in $B$ decays and in establishing factorisation of the decay matrix elements.
 In exclusive decays, one also needs the decay form factors and a lot of theoretical progress has been made using the lattice QCD \cite{Aoki:2016frl} and QCD sum rule techniques
 \cite{Ali:1993vd,Colangelo:2000dp,Ball:2004ye,Straub:2015ica}, often complementing each other,
as they work best in the opposite $q^2$-ranges. It is this continued progress in QCD calculational framework,
which has  taken us to the level of sophistication required  to match the experimental advances.

In this paper, I review what, in my view, are some of the key measurements in the radiative, semileptonic and
leptonic rare $B$-decays and confront them with the SM-based calculations, carried out with the
theoretical tools just mentioned. However, this is not a comprehensive review of this subject, but the hope
is that the choice of topics reflects both the goals achieved in explaining some landmark measurements and focus on 
open issues. In section 2, I review the inclusive and some exclusive radiate rare $B$-decays.There are no
burning issues in this area - at least not yet. In section 3, the corresponding inclusive and exclusive 
semileptonic decays are taken up. Again, there are no open issues in the inclusive semileptonic decays, but
experimental precision is limited currently, which is bound to improve significantly at Belle II. There are, however,
a lot of open issues in the exclusive semileptonic decays, in particular in $R_K$, the ratio of the
decay widths for $B \to K \mu^+ \mu^-$ and $B \to K e^+ e^-$, hinting at the possible breakdown of lepton universality,
the linchpin of the SM, reviving the interest in low-mass leptoquarks. One should also mention here
similar issues in tree-level semileptonic decays, such as $R^{\tau/\ell}_D$ and $R^{\tau/\ell}_{D^*}$,
the ratios involving the decays $B \to D^{(*)} \tau \nu_\tau$ and $B \to D^{(*)} \ell \nu_\ell$ ($\ell= e, \mu$).
There are also other dissenting areas, which go by the name of $P_5^\prime$-anomaly, which is a certain coefficient
in the angular description of the decay $B \to K^* \mu^+ \mu^-$, which presumably need 
 a better theoretical (read QCD) description than is available at present. They are discussed in section 3.
 In section 4, we discuss the CKM-suppressed $b \to d \ell^+ \ell^-$ decays, which is a new experimental
 frontier initiated by the LHCb measurements of the branching fraction and the dimuon invariant mass distribution
 in the decay $B^\pm \to \pi^\pm \mu^+ \mu^-$.  Finally, the rarest $B$- and $B_s$-decays measured so far,
 $B \to \mu^+ \mu^-$  and $B_s \to \mu^+ \mu^-$,  are taken up in section 5. Current measurements also
 show some (mild) deviations in their branching ratios versus the SM. A representative global fit of the data on the semileptonic  and leptonic rare $B$-decays in terms of the Wilson coefficients from possible new physics is shown in section 6. Some concluding remarks are  made in section 7.

\section{Rare  Radiative $B$-decays in the SM and Experiments} 
\begin{figure}
\includegraphics[width=2.0in]{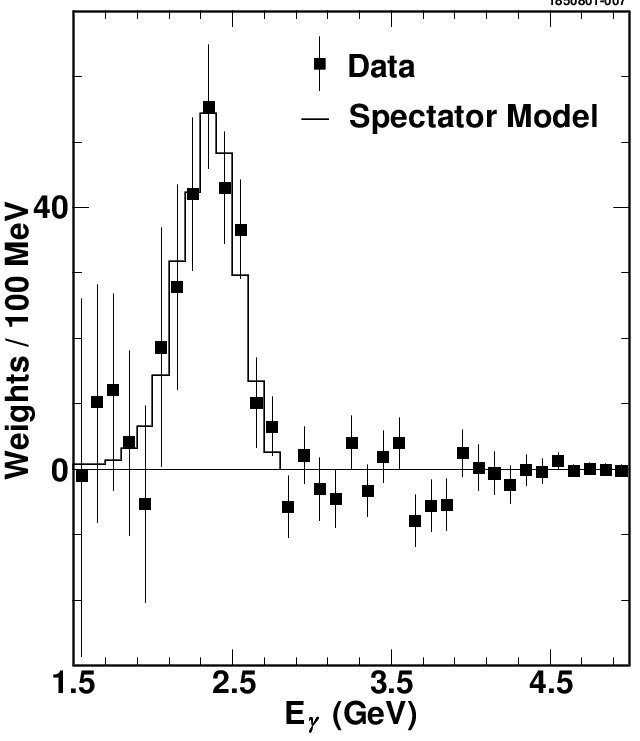}
\includegraphics[width=2.35in]{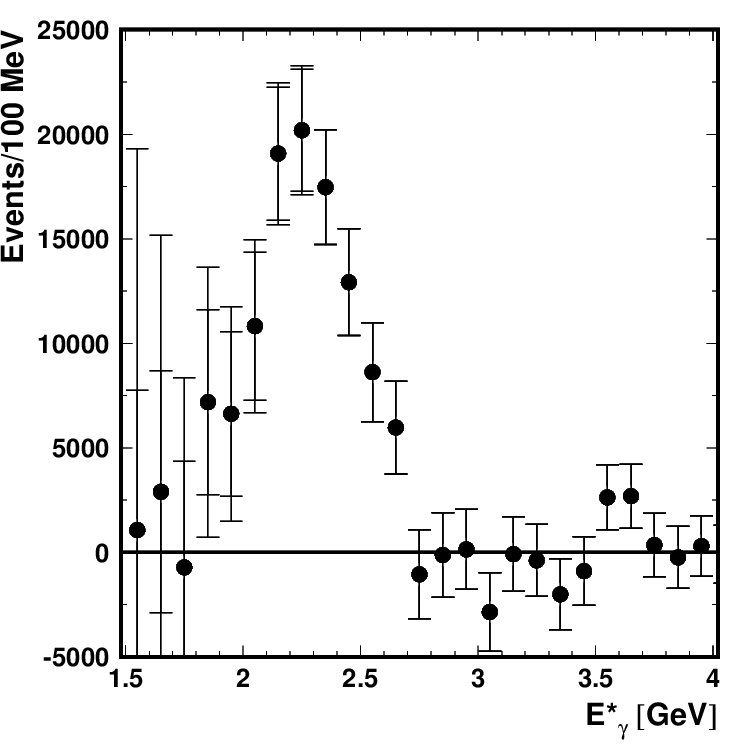}
\caption{Photon energy spectrum in the inclusive decay $B \to X_s \gamma$
measured by CLEO (left frame) \cite{Chen:2001fja} and Belle (right frame) \cite{Koppenburg:2004fz}.}
\label{fig:belle-celo-bsgama}
\end{figure}

In 1993, the CLEO collaboration at the Cornell $e^+e^-$ collider measured the decay
 $B \to K^* \gamma$ \cite{Ammar:1993sh}, 
initiating the field of rare $B$-decays,  followed two years later by the measurement of
the inclusive decay  $B \to X_s \gamma$  \cite{Alam:1994aw}.
The branching ratio $ {\cal}(B \to K^* \gamma)  = (4.5 \pm 1.5 \pm 0.9)\times 10^{-5}$ was in agreement
with the SM estimates, but theoretical uncertainty was large. Measuring the Inclusive process
 $B \to X_s \gamma$ was challenging, but as the
 photon energy spectrum in this process was already
calculated in 1990 by Christoph Greub and me\cite{Ali:1990vp}, this came in handy for the
 CLEO measurements \cite{Chen:2001fja}
  shown in Fig. \ref{fig:belle-celo-bsgama} (left frame) and compared with the  theoretical prediction \cite{Ali:1990vp}, 
Since then, a lot of experimental and theoretical effort has gone in the precise measurements and in performing
higher order perturbative and non-perturbative calculations. As a consequence, 
$ B \to X_s \gamma$ has now become the standard candle of FCNC processes, with the measured
branching ratio and the precise higher order SM-based calculation providing valuable constraints on
the parameters of BSM physics. The impact of the $B$-factories on this measurement can be judged by the scale in Fig. \ref{fig:belle-celo-bsgama} (right frame), 
 which is due to the Belle collaboration \cite{Koppenburg:2004fz}.

The next frontier of rare $B$-decays involves the so-called electroweak penguins, which govern
the decays of the inclusive processes $ B \to (X_s , X_d) \ell^+ \ell^-$ and the exclusive decays
such as $B \to (K,K^*, \pi) \ell^+ \ell^-$. These processes have rather small branching ratios and hence
they were first measured at the $B$-factories. Inclusive decays remain their domain, but
experiments at the LHC, in particular, LHCb, are now at the forefront of exclusive semileptonic decays.
Apart from these, also the leptonic $B$-decays $  B_s \to \mu^+\mu^-$ and
 $ B_d \to \mu^+\mu^-$ have been measured at the LHC.  

I will review some of the  key measurements
and the theory relevant for their interpretation. This description is anything but comprehensive, for which I refer to some recent excellent references \cite{Blake:2016olu,Descotes-Genon:2015uva,Koppenburg:2016rji,Bevan:2014iga} 
and resources, such as HFAG \cite{Amhis:2014hma} and  FLAG  \cite{Aoki:2016frl}.

\subsection{Inclusive decays $B \to X_s \gamma$ at NNLO in the SM}
\begin{figure}
\centerline{\includegraphics[width=3.5in]{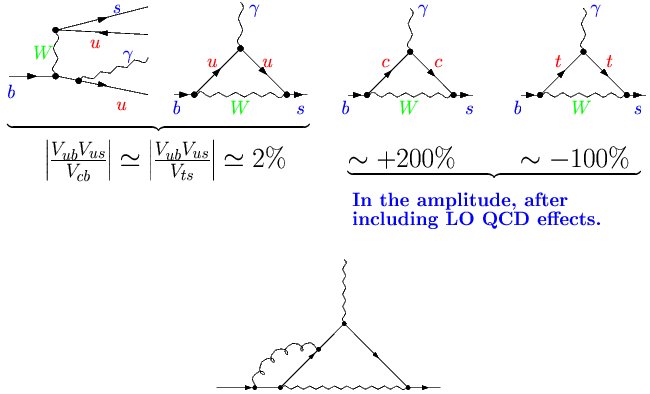}}
\caption{Examples of  the leading electroweak diagrams for  $B \to X_s \gamma$ from the up,
charm, and top quarks. A  diagram involving a gluon exchange
is shown in the lower figure.}
\label{fig:leading-bsgamma-diag}
\end{figure}

The leading order diagrams for the decay $b \to s \gamma$ are shown are shown in 
Fig. \ref{fig:leading-bsgamma-diag},  including also the tree diagram for $b \to u \bar{u} s \gamma$, which yields a
soft photon. The first two diagrams are anyway suppressed due to the CKM matrix elements, as indicated.
The  charm- and top- quark contributions enter with opposite signs, and the relative contributions
indicted are after including the leading order (in $\alpha_s$)
QCD effects. A typical diagram depicting perturbative QCD corrections due to the exchange of a gluon
is also shown. 

The QCD logarithms $\alpha_s \ln M_W^2/m_b^2$ enhance the branching ratio ${\cal B}(B \to X_s \gamma)$
by more than a factor 2, and hence such logs have to be resummed. This is done using an effective field theory
approach, obtained by integrating out the top quark and the $W^\pm$ bosons. Keeping terms up to dimension-6,
the effective Lagrangian for $B \to X_s \gamma$ and $B \to X_s \ell^+ \ell^-$  reads as follows:
 \begin{eqnarray}
{\cal L} \;= \;\; {\cal L}_{QCD \times QED}(q,l) 
\;\;+ \frac{4 G_F}{\sqrt{2}} V_{ts}^* V_{tb} \sum_{i=1}^{10} C_i(\mu) O_i \nonumber
\end{eqnarray}
($q=u,d,s,c,b$,~~$l = e,\mu, \tau$)
\begin{eqnarray}
O_i = \left\{ \begin{array}{lll}
(\bar{s} \Gamma_i c)(\bar{c} \Gamma'_i b), 
                         & i=1, (2), & C_i(m_b) \sim -0.26 ~(1.02)\\[4mm]
(\bar{s} \Gamma_i b) {\Sigma}_q (\bar{q} \Gamma'_i q), 
                         & i=3,4,5,6,~ & |C_i(m_b)| < 0.08\\[4mm]
\frac{e m_b}{16 \pi^2} \bar{s}_L \sigma^{\mu \nu} b_R F_{\mu \nu}, 
                         & i=7, & C_7(m_b) \sim -0.3\\[4mm]
\frac{g m_b}{16 \pi^2} \bar{s}_L \sigma^{\mu \nu} T^a b_R G^a_{\mu \nu}, 
                         & i=8, & C_8(m_b) \sim -0.16\\[4mm]
\frac{e^2}{16 \pi^2} (\bar{s}_L \gamma_{\mu} b_L) (\bar{l} 
\gamma^{\mu} {(\gamma_5)} l),
                         & i=9,{(10)} & C_i(m_b) \sim 4.27~ (-4.2) 
\end{array} \right.  
\nonumber
\end{eqnarray}
Here, $G_F$ is the Fermi coupling constant, $V_{ij}$ are the CKM matrix elements, $O_i$ are the
four-Fermi and dipole operators, and $C_i(\mu)$ are the Wilson coefficients, evaluated
at the scale $\mu$, which is taken typically as $\mu=m_b$, and their values in the NNLO accuracy
are given above for $\mu=4.8$ GeV.  Variations due to a different choice
of $\mu$ and uncertainties from  the upper scale-setting $m_t/2 \leq  \mu_0 \leq 2 m_t$ can be seen elsewhere
\cite{Blake:2016olu}.

There are three essential steps of the calculation:
\vspace*{-3mm}
\begin{itemize}
\item \underline{Matching}:~~Evaluating 
 $C_i(\mu_0)$ at  $\mu_0 \sim M_W$ by requiring the equality of the SM and 
the effective theory Green functions.

\item \underline{Mixing}:~~ Deriving the effective theory renormalisation group equation (RGE) 
and evolving $C_i(\mu)$ from $\mu_0$ to $\mu_b \sim m_b$. 

 \item \underline{Matrix elements}:~~ Evaluating the on-shell 
amplitudes at  $\mu_b \sim m_b$.
\end{itemize}
\vspace*{-3mm}
All three steps have been improved in perturbation theory and now include the next-to-next-to-leading order effects (NLLO),
i.e., contributions up to $O(\alpha_s^2(m_b))$. A monumental theoretical effort stretched well over a decade 
with the participation of a large number of theorists underlies the current theoretical precision of the branching
ratio. The result is usually quoted for a threshold photon energy to avoid experimental background from other
Bremsstrahlung processes. For  the decay  with $E_\gamma > 1.6$ GeV in the rest frame of the $B$ meson,
the result at NNLO accuracy is \cite{Misiak:2015xwa,Czakon:2015exa}
\begin{equation}
{\cal B}(B \to X_s \gamma)= (3.36 \pm 0.23) \times 10^{-4},
\end{equation}
where the dominant SM uncertainty is non-perturbative \cite{Benzke:2010js}. This is
to be compared with the current experimental average of the same \cite{Amhis:2014hma}
\begin{equation}
{\cal B}(B \to X_s \gamma)= (3.43 \pm 0.21 \pm 0.07) \times 10^{-4},
\end{equation}
where the first error is statistical and the second systematic, yielding a ratio $1.02 \pm 0.08$,
providing a test of the SM to an accuracy better than 10\%.

The CKM-suppressed decay $B \to X_d \gamma$ has also been calculated in the NNLO precision.
The result for $E_\gamma > 1.6$ GeV is  \cite{Misiak:2015xwa}

\begin{equation}
{\cal B}(B \to X_d \gamma)= (1.73 ^{+0.12}_{-0.22}) \times 10^{-5}.
\end{equation}
This will be measured precisely at Belle II. The constraints on the CP asymmetry are not very restrictive, but the
current measurements are in agreement with the SM expectation. For further details, see HFAG \cite{Amhis:2014hma}.

\subsection{Bounds on the charged Higgs mass from ${\cal B}(B \to X_s \gamma)$}
\begin{figure}
\centerline{\includegraphics[width=3.5in]{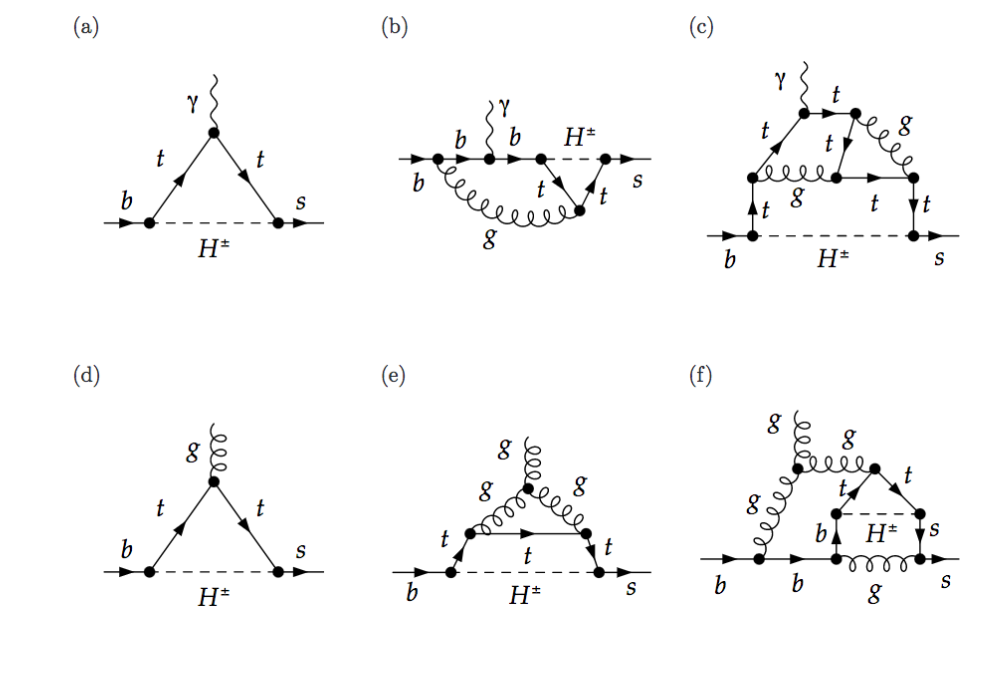}}
\caption{Sample Feynman diagrams that matter for  $B \to X_s \gamma$ in the 2HDM \cite{Hermann:2012fc}.
 $H^\pm$ denotes a charged Higgs. }
\label{fig:2HDM-bsgamma-diag}
\end{figure}

\begin{figure}
\centerline{\includegraphics[width=3.5in]{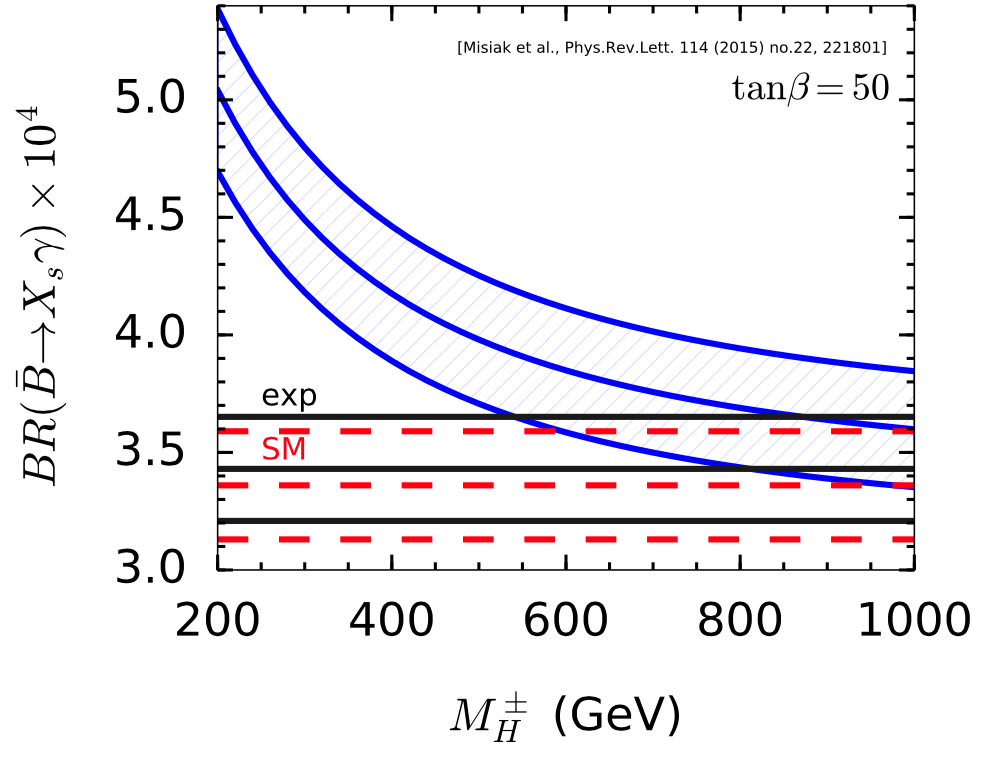}}
\caption{Constraints on the charged Higgs mass $m_H^\pm$ from
  ${\cal B}(B \to X_s \gamma)$ in the 2HDM \cite{Misiak:2015xwa,Hermann:2012fc}. Measured branching ratio (exp) and the SM
  estimates are also shown. The curves demarcate the central values and  $\pm 1\sigma$ errors.}
\label{fig:2HDM-Steinhauser}
\end{figure}

As the agreement between the SM and data is excellent, the decay rate for $B \to X_s \gamma$ provides constraints on
the parameters of the BSM theories, such as supersymmetry and the 2 Higgs-doublet models
(2HDM). In calculating the BSM effects, depending on the model,  the SM operator basis may have to be enlarged,
but in many cases one anticipates additive contributions to the Wilson coefficients in the SM basis. In the context
of  $ B \to X_s \gamma$, it is  customary to encode the BSM effects in the Wilson coefficients of the
dipole operators $C_7(\mu)$ and  $C_8(\mu)$, and the constraints from the branching ratio
on the additive coefficients  $\Delta C_7$ and  $\Delta C_8$  then takes
the numerical form  \cite{Misiak:2015xwa}
\begin{equation}
{\cal B}(B \to X_s \gamma) \times 10^4 = (3.36 \pm 0.23)  -8.22 \Delta C_7 -1.99 \Delta C_8.
\end{equation}
To sample the kind of constraints that can be derived on the parameters of the BSM models, the 2HDM is a
good case, as the branching ratio for the decay $B \to X_s \gamma$ in this model has been derived to the same theoretical accuracy \cite{Hermann:2012fc}.The Lagrangian for the 2HDM is
\begin{equation}
 {\cal L}_{H^+}= (2\sqrt{2} G_F)^{1/2}\Sigma_{i,j=1}^{3} \bar{u}_i (A_u m_{u_i}
V_{ij}P_L - A_d m_{d_j}V_{ij} P_R) d_j H^* + h.c.,
\end{equation}
where $V_{ij}$ are the CKM matrix elements and 
$P_{L/R}=(1 \mp \gamma_5)/2 $. The 2HDM contributions to the Wilson coefficients are proportional to
$A_iA_j^*$, representing the contributions from the up-type $ A_u$ and down-type $ A_d $ quarks. They
are defined in terms of the ratio of the vacuum expectation values, called $\tan \beta$, and are model dependent.
\begin{itemize}
\item 2HDM of type-I:  $A_u=A_d=\frac{1}{\tan \beta},$
\item 2HDM of type-II:  $A_u=-1/A_d=\frac{1}{\tan \beta}$.
\end{itemize}
Examples of Feynman diagrams contributing to $B \to X_s \gamma$ in the 2HDM are shown in 
Fig. \ref{fig:2HDM-bsgamma-diag}.
Apart from  $\tan \beta$, the other parameter of the 2HDM is the mass of the charged Higgs $M_H^\pm$.
As  ${\cal B}(B \to X_s \gamma)$
becomes insensitive to $\tan \beta$ for larger values,  $\tan \beta>2$ , the 2HDM contribution depends essentially on $M_H^\pm$. The current measurements and the SM estimates then provide constraints on  $M_H^\pm$,
as shown in Fig. \ref{fig:2HDM-Steinhauser}, \footnote{I thank Matthias Steinhauser for providing this figure.}
 updated using  \cite{Misiak:2015xwa,Hermann:2012fc},
 yielding \cite{Misiak:2015xwa}
$M_H^\pm > 480$ GeV (\@ 90\% C.L.) and  $M_H^\pm > 358$ GeV (\@ 99\% C.L.) .
 These constraints are
competitive  to the direct searches of the $H^\pm$ at the LHC. %

\subsection{Exclusive radiative rare $B$ decays}
Exclusive radiative decays, such as $B \to V \gamma$ ($V=K^*, \rho, \omega $) and
$B_s \to \phi \gamma$,  have been well-measured at the $B$ factories. In addition, they
offer the possibility of measuring  CP- and isospin asymmetries, a topic I will not discuss here. Theoretically,
exclusive decays are more
challenging, as they require the knowledge of the form factors at $q^2=0$, which can not be
calculated directly using Lattice QCD. However, light-cone QCD sum rules \cite{Ball:2004ye,Straub:2015ica} 
also do a good job for
calculating heavy $\to$ light form factors at low-$q^2$. In addition, the matrix elements require gluonic exchanges between the spectator quark and the active quarks (spectator-scattering),
introducing intermediate scales in the decay rates. Also long-distance 
effects generated by the four-quark operators with charm quarks are present and are calculable in
limited regions \cite{Khodjamirian:2010vf}. 
Thus,  exclusive decays are  theoretically not as precise as the inclusive decay $B \to X_s \gamma$. However, techniques embedded in HQET and SCET have led to the
factorisation of the decay matrix elements into perturbatively calculable (hard) and non-perturbative
(soft) parts, akin to the deep inelastic scattering processes. These factorisation-based approaches
are the main work-horse in this field. Renormalisation group (RG) methods then allow to
sum up large logarithms, and this program has been carried out to a high accuracy. 

A detailed discussion of the various techniques requires a thorough review, which can't be carried out here.
I will confine myself by pointing to some key references, beginning from the QCD factorisation approach,
pioneered by Beneke, Buchalla, Neubert and Sachrajda \cite{Beneke:1999br}, which has been applied to
the radiative decays $B \to (K^*, \rho, \omega) \gamma$ 
\cite{Beneke:2001at,Ali:2001ez,Bosch:2001gv,Beneke:2004dp}. Another theoretical framework, called
pQCD \cite{Li:1994iu,Keum:2000wi},   has also been put to use in these decays \cite{Keum:2004is,Lu:2005yz}.
The SCET-based methods have also been harnessed \cite{Chay:2003kb,Becher:2005fg}.
The advantage of SCET is that it allows for an unambiguous separation of the scales, and an operator  definition
of each object in the factorisation formula can be given.  Following the QCD factorisation
approach, a factorisation formula  for the $B \to V \gamma$ matrix element can be written in SCET as well
\begin{equation}
\langle V \gamma \vert Q_i \vert B \rangle = \Delta_i C^{A} \xi_{V_\perp}
+ \frac{ \sqrt{m_B} F f_{V_\perp}} {4} \int dw du\; \phi_+^{B} (w)  \phi_\perp^{V}(u)\; t_i^{II},
\label{eq:scet-fact}
\end{equation}
where $F$ and $ f_{V_\perp} $ are meson decay constants; $\phi_+^{B} (w)$ and $\phi_\perp^{V}(u) $ are the
light-cone distribution amplitudes for the $B$- and $V$-meson, respectively.
 The SCET form factor $ \xi_{V_\perp} $ is
related to the QCD form factor through perturbative and power corrections, and the perturbative hard QCD
kernels are the coefficients $ \Delta_i C^{A} $ and $t_i^{II} $. They are known to complete NLO accuracy in
RG-improved perturbation theory \cite{Becher:2005fg}.

 The factorisation formula (\ref{eq:scet-fact}) has been calculated to NNLO accuracy  in SCET \cite{Ali:2007sj} (except for the NNLO corrections from the spectator scattering). As far as the
decays $B \to K^* \gamma$ and $B_s \to \phi \gamma$ are concerned, the partial NNLO  theory 
is still the state-of-the-art.
Their  branching ratios  as well as the ratio of the decay rates 
$ {\cal B}(B_s \to \phi \gamma/{\cal B}(B \to K^* \gamma)$ are given in Table \ref{tab:exclusive-rad-decays},
together with the current experimental averages \cite{Amhis:2014hma}. The corresponding calculations for the CKM-suppressed
decays $ B \to (\rho,\omega)  \gamma$ are not yet available to the desired theoretical accuracy, due to the annihilation contributions, for which,
to the best of my knowledge, no factorisation theorem of the kind discussed above has been proven.
The results from a QCD-Factorisation based approach \cite{Ali:2001ez} 
for $ B \to \rho \gamma$ are also
given in Table  \ref{tab:exclusive-rad-decays}  and compared with the data.  The exclusive decay rates
shown are in agreement with the experimental measurements, though theoretical precision is not better than 20\%.   Obviously, there is need for a better theoretical description, more so as Belle II will measure the radiative decays with greatly improved precision. I will skip a discussion of the isospin and CP asymmetries in these decays,
as  the current experimental bounds  \cite{Amhis:2014hma} are not yet probing the SM in these observables.

\begin{table}
\begin{center}
\tbl{Measurements [HFAG 2014] \cite{Amhis:2014hma} and SM-based estimates  of 
 ${\cal B}(B \to ( K^*, \rho) \gamma)$ and ${\cal B}(B_s \to  \phi  \gamma)$
 in units of $ 10^{-5}$, and the ratio ${\cal B}(B_s \to  \phi  \gamma)/{\cal B}(B^0 \to  K^{*0} \gamma)$. }
{\begin{tabular}{| l | l | l |}\hline
Decay Mode &  Expt.~(HFAG)
& Theory (SM) \gh \\ \hline
$B^0 \to K^{*0} \gamma$&$ 4.33 \pm 0.15$
&   $4.6 \pm 1.4 $  \gh \\
 \hline
$B^+\to K^{*+}  \gamma $&$4.21 \pm 0.18$ &  $ 4.3 \pm 1.4$ \gh \\
\hline
$B_s\to \phi \gamma $&$ 3.59 \pm 0.36 $ & $ 4.3 \pm 1.4 $ \gh \\ \hline
$ B_s\to \phi \gamma/ B^0 \to K^{*0} \gamma $ & $0.81 \pm 0.08$ &  $1.0 \pm 0.2$ \gh
 \\ \hline
$B^0 \to \rho^0  \gamma  $ & $0.86^{+0.15}_{-0.14}$ &  $  0.65 \pm 0.12$ \gh \\ 
\hline
$B^+ \to \rho^+  \gamma  $ & $0.98  \pm 0.25 $ &  $ 1.37 \pm 0.26$ \gh \\
\hline
\label{tab:exclusive-rad-decays}
\end{tabular}}
\end{center}
\end{table}

\section{Semileptonic $b \to s$ decays $B \to (X_s , K, K^*) \ell^+ \ell^-$}

There are two $b \to s$ semileptonic operators in SM:
\begin{equation}
O_i = \frac{e^2}{16 \pi^2} (\bar{s}_L \gamma_{\mu} b_L) 
       (\bar{l} \gamma^{\mu} (\gamma_5) l), \hspace{1.5cm} i=9,(10)
\nonumber
\end{equation}
Their Wilson Coefficients have the following perturbative expansion:
\begin{eqnarray}
C_9(\mu) &=& \frac{4 \pi}{\alpha_s(\mu)} C_9^{(-1)}(\mu) 
        ~+~ C_9^{(0)}(\mu) ~+~ \frac{\alpha_s(\mu)}{4 \pi} C_9^{(1)}(\mu) 
~+~ ...\nonumber\\ 
C_{10} &=&  C_{10}^{(0)} ~+~\frac{\alpha_s(M_W)}{4 \pi} C_{10}^{(1)} ~+~ ...
\nonumber
\end{eqnarray}
 The term 
$C_9^{(-1)}(\mu)$ reproduces the electroweak
logarithm that originates from the photonic penguins with charm quark
loops, shown below \cite{Ghinculov:2003qd}.
\centerline{\includegraphics[width=3.5cm,angle=0]{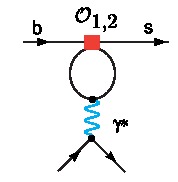}}
The first two terms in the perturbative expansion of $C_9(m_b)$ are
 \begin{eqnarray}
&& C_9^{(0)}(m_b) \simeq ~2.2; ~~\frac{4 \pi}{\alpha_s(m_b)} \,C_9^{(-1)}(m_b) =
\frac{4}{9}\, \ln \frac{M_W^2}{m_b^2} + {\cal O}(\alpha_s) \simeq~2. 
\nonumber
\end{eqnarray}
 As they  are very similar in magnitude,
one needs to calculate the NNLO contribution to get reliable estimates of the decay rate.
In addition, leading power corrections in  $1/m_c$ and $1/m_b$ are required.

\subsection{Inclusive semileptonic decays $B \to X_s  \ell^+ \ell^-$}
\begin{figure}
\centerline{\includegraphics[width=3.5in]{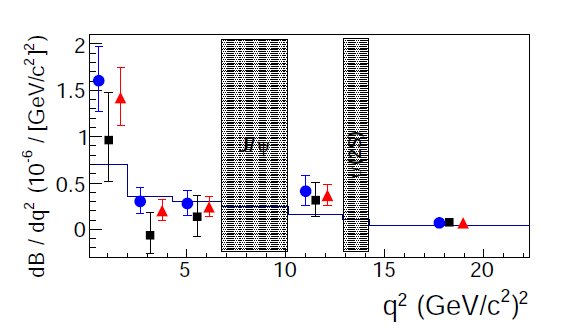}}
\centerline{\includegraphics[width=3.5in]{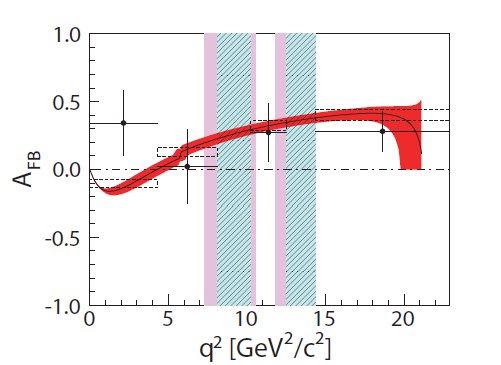}}
\caption{Dilepton invariant mass Distribution measured by BaBar \cite{Lees:2013nxa}  (upper frame) and the Forward-backward Asymmetry $A_{\rm FB}$ measured by Belle \cite{Sato:2014pjr} (lower frame) in $B \to X_s \ell^+ \ell^-$.
The curve(above) and the band (below)  are the SM expectations, discussed in the text.}
\label{fig:belle-babar-bsll}
\end{figure}

A lot of theoretical effort has gone into calculating the perturbative 
QCD NNLO , electromagnetic logarithms and power corrections
 \cite{Ghinculov:2003qd,Asatryan:2001zw,Ali:2002jg,Huber:2005ig,Huber:2007vv}. 
The B-factory experiments
Babar and Belle have measured the dilepton invariant mass spectrum  $d{\cal B}(B \to X_s \ell^+ \ell^-)/dq^2$
practically in the entire kinematic region
and have also measured the so-called Forward-backward lepton asymmetry $A_{FB}(q^2)$ \cite{Ali:1991is}. They are shown in
 Fig. \ref{fig:belle-babar-bsll}, and compared with the SM-based theoretical calculations. 
 Note that a cut of $q^2 > 0.2$ GeV$^2$ on the dilepton invariant  squared mass is used.
 As seen in these
 figures, two resonant regions near $q^2=M_{J/\psi}^2$ and $q^2=M_{J/\psi^\prime}^2$ have to be excluded
 when comparing with the short-distance contribution. 
  They make up what is called the long-distance contribution from the processes
 $B \to X_s + (J/\psi, J/\psi^\prime) \to X_s + \ell^+ \ell^-$, whose dynamics is determined by the hadronic matrix
 elements of the operators $O_1$ and $O_2$. They have also been calculated via 
 a dispersion relation \cite{Kruger:1996cv} and data on the measured quantity 
 $R_{\rm had} (s)= \sigma (e^+ e^- \to {\rm hadrons})/\sigma (e^+ e^- \to \mu^+ \mu ^-)$, and in some
 analyses are also included. 
 As the (short-distance)  contribution is expected
 to be a smooth function of $q^2$, one uses the perturbative distributions in interpolating  these
 regions as well. The experimental distributions are in agreement with the SM, including also the zero point
 of $A_{FB}(q^2)$, which is a sensitive function of the ratio of the two Wilson coefficients $C_9$ and $C_{10}$.
  
The branching ratio for the inclusive decay  $B \to X_s \ell^+ \ell^-$  with a lower cut on the
dilepton invariant mass $q^2 > 0.2~ {\rm GeV}^2$ at NNLO accuracy is \cite{Ali:2002jg}
\begin{equation}
{\cal B}(B \to X_s \ell^+ \ell^-)= (4.2 \pm 0.7) \times 10^{-6},
\end{equation}
to be compared with the current experimental average of the same \cite{Amhis:2014hma}
\begin{equation}
{\cal B}(B \to X_s  \ell^+ \ell^-)= (3.66^{+0.76}_{-0.77}) \times 10^{-6}.
\end{equation}
The two agree within theoretical and experimental errors. The experimental cuts which are imposed to
remove the $J/\psi$ and $\psi^\prime$ resonant regions are indicated in Fig. \ref{fig:belle-babar-bsll}.
 The effect of logarithmic QED corrections becomes
important for more restrictive cuts on $q^2$, and they have been worked out for different choices of the
$q^2$-range in a recent paper \cite{Huber:2015sra}.

\subsection{Exclusive Decays $B \to (K,K^*) \ell^+ \ell^-$}

The  $B \to K$ and  $B \to K^*$ transitions involve the following weak currents:
\begin{equation}
 \Gamma_\mu^1=\bar{s}\gamma_\mu (1-\gamma_5) b, ~~
\Gamma_\mu^2=\bar{s}\sigma_{\mu\nu}q^\nu (1+\gamma_5) b. 
\end{equation}
Their matrix elements involve altogether 10 non-perturbative $q^2$-dependent functions (form factors),
denoted by the following functions: \footnote{As we will also discuss later the decays $ B \to \pi \ell^+ \ell^- $,
we distinguish the $B \to K$  and $B \to \pi$ form factors by a superscript.}
\begin{eqnarray}
\langle K \vert \Gamma_\mu^1 \vert B \rangle && \supset f^K_+(q^2),
f^K_-(q^2). \nonumber \\
\langle K \vert \Gamma_\mu^2 \vert B \rangle && \supset f_T^K(q^2). \nonumber \\
\langle K^* \vert \Gamma_\mu^1 \vert B \rangle  && \supset V(q^2), 
A_1(q^2),A_2(q^2),A_3(q^2). \nonumber \\
\langle K^* \vert \Gamma_\mu^2 \vert B \rangle  && \supset 
T_1(q^2),T_2(q^2),T_3(q^2).\nonumber 
\end{eqnarray}
 Data on $B \to K^* \gamma$ provides normalisation of $T_1(0)=T_2(0)
\simeq 0.28$. These form factors have been calculated using a number of non-perturbative
techniques, in particular the QCD sum rules  \cite{Ball:2004ye,Bharucha:2010im} and Lattice QCD
 \cite{Dalgic:2006dt,Bouchard:2013zda,Bailey:2015nbd}.
 They are complementary to
each other, as the former are reliable in the low-$q^2$ domain and the latter can calculate
only for large-$q^2$. They are usually combined to get reliable profiles of the form factors in
the entire $q^2$ domain. However, heavy quark symmetry  allows to reduce the number of independent form
factors from 10 to 3 in low-$q^2$ domain $(q^2/m_b^2 \ll 1)$. 
Symmetry-breaking corrections have been evaluated \cite{Beneke:2000wa}. 
The decay rate,  dilepton invariant mass distribution and the
Forward-backward asymmetry  in the low-$q^2$ region  have been calculated for $B \to K^* \ell^+ \ell^-$
 using the SCET formalism  \cite{Ali:2006ew}.
Current measurements of the branching ratios in the inclusive and exclusive semileptonic decays
 involving $b \to s$ transition are summarised in Table \ref{tab:exclusive-decays} and  compared with the
 corresponding SM estimates. 
 The inclusive measurements and the SM rates include a cut on the dilepton invariant mass 
$M_{\ell^+\ell^-} > 0.2$ GeV.
 They are in agreement with each other, though precision is
 currently limited due to the imprecise knowledge of the form factors.
 
 \subsection{Current tests of lepton universality in semileptonic $B$-decays}
  Currently, a number of measurements in $B$ decays suggests a breakdown of the lepton $(e, \mu, \tau)$
  universality in semileptonic processes. In the SM, gauge bosons couple with equal strength to all three
  leptons and the couplings of the Higgs to a pair of charged lepton is proportional to the charged lepton mass,
  which are negligibly small for $\ell^+ \ell^-= e^+ e^-, \mu^+ \mu^-$. Hence, if the lepton non-universality is
  experimentally established, it would be a fatal blow to the SM. 
  
   We briefly summarise the experimental situation starting from the decay $ B^\pm \to K^\pm \ell^+  \ell^- $,
   whose decay rates were discussed earlier.
Theoretical accuracy is vastly improved if instead of the absolute rates, ratios of the decay rates
are computed. 
  Data on the decays involving  $K^{(*)} \tau^+ \tau^-$ is currently
sparse, but first measurements of the ratios involving  the final states 
$K^{(*)} \mu^+ \mu^-$ and $ K ^{(*)} e^+ e^- $ are available. In particular, a 2.6$\sigma$ deviation from the
$e$-$\mu$ universality is reported by the LHCb collaboration in the ratio involving $B^\pm \to K^\pm \mu^+ \mu^-$
and $B^\pm \to K^\pm  e^+ e^-$ measured in the low-$q^2$ region,
which can be calculated rather accurately.
 In the interval $ 1 \leq q^2 \leq 6$ GeV$^2$, LHCb finds \cite{Aaij:2014ora}
\begin{equation}
R_K \equiv \frac{\Gamma(B^\pm \to K^\pm \mu^+ \mu^-)} {\Gamma(B^\pm  \to K^\pm  e^+ e^-)} 
= 0.745^{+0.090}_{-0.074} ({\rm stat}) \pm 0.035 ({\rm syst}).
\label{eq:RK-anomaly}
\end{equation}
This ratio in the SM is close to 1 to a very high accuracy \cite{Bobeth:2007dw} over the entire $q^2$ region measured by the LHCb. Thus, the measurement  in (\ref{eq:RK-anomaly}) amounts to
 about $2.6\sigma$  deviation from the SM. Several BSM scenarios have been proposed to account for the
 $R_K$ anomaly, discussed below,  including a $Z^\prime$-extension of the SM \cite{Chiang:2016qov}.
 It should, however, be noted that the currently measured
 branching ratios 
  ${\cal B}(B^\pm \to K^\pm e^+ e^-)= (1.56^{+ 0.19 + 0.06}_{-0.15 -0.4})\times 10^{-7} $ and
  ${\cal B}(B^\pm \to K^\pm \mu^+ \mu^-)= (1.20 \pm 0.09 \pm 0.07)\times 10^{-7} $
    are also lower than the SM estimates  ${\cal B}^{\rm SM}(B^\pm \to K^\pm e^+ e^-)=
 {\cal B}^{\rm SM} (B^\pm \to K^\pm \mu^+ \mu^-)= (1.75^{+0.60}_{-0.29}) \times 10^{-7}$, and the experimental
 error on the  ${\cal B}(B^\pm \to K^\pm e^+ e^-) $ is twice as large. One has to also factor in that the electrons
 radiate very profusely (compared to the muons) and implementing the radiative corrections in hadronic machines
 is anything but straight forward. 
 In coming years, this and similar ratios, which can also be calculated to high accuracy,  will be measured with greatly
improved precision at the LHC and Belle II. 

The other place where lepton non-universality is reported is in the ratios of the decays
$B \to D^{(*)} \tau \nu_\tau$ and $B \to D^{(*)} \ell \nu_\ell$. Defining 
\begin{equation}
R^{\tau/\ell}_{D^{(*)}} \equiv \frac{{\cal B}(B \to D^{(*)} \tau \nu_\tau)/ {\cal B} ^{\rm SM}(B \to D^{(*)} \tau \nu_\tau) }
{{\cal B}(B \to D^{(*)} \ell \nu_\ell)/ {\cal B} ^{\rm SM}(B \to D^{(*)} \ell \nu_\ell)},
\label{eq:RDDS-tau}
\end{equation}
the current averages of the BaBar, Belle, and the LHCb data are \cite{Amhis:2014hma}:
\begin{equation}
R^{\tau/\ell}_D= 1.37 \pm 0.17;\hspace{3mm } R^{\tau/\ell}_{D^*}= 1.28 \pm 0.08.
\label{eq:RDDS-tau-expt}
\end{equation}
This amounts to about $3.9\sigma$ deviation from the $\tau/\ell $ ($\ell= e, \mu $) universality. Interestingly, this
happens in a tree-level charged current process. If confirmed experimentally, this would call for a 
drastic contribution to an effective four-Fermi $LL$ operator
 $( \bar{c} \gamma_\mu b_L)(\tau_L \gamma_\mu \nu_L  )$. It is then conceivable that the 
 non-universality in $R_K$ (which is a loop-induced $b \to s$ process) ia also due to an $LL$ operator
 $( \bar{s} \gamma_\mu b_L)(\bar{\mu}_L \gamma_\mu \mu_L )$. Several suggestions along these lines
 involving a leptoquark have been made \cite{Hiller:2014yaa,Bauer:2015knc,Barbieri:2015yvd}. It is 
 worth recalling that leptoquarks were introduced by Pati and Salam in 1973 in an attempt to unify
 leptons and quarks in $SU(4)$ \cite{Pati:1973uk,Pati:1974yy}. The lepton non-universality in $B$ decays
 has revived the interest in theories with low-mass leptoquarks, discussed recently
 in a comprehensive work on this topic \cite{Dorsner:2016wpm}.

\begin{table}
\begin{center}
\tbl{Measurements [PDG 2014] and SM-based estimates \cite{Ali:2002jg} of the
branching ratios ${\cal B}(B \to (X_s, K, K^*) \ell^+ \ell^-)$  in units of $ 10^{-6}$}
{\begin{tabular}{| l | l | l |}\hline
Decay Mode &  Expt.~(BELLE \& BABAR)
& Theory (SM) \gh \\ \hline
$B\to K\ell^+\ell^-$&$ 0.48\pm 0.04$
&   $0.35 \pm 0.12 $  \gh \\
 \hline
$B\to K^*e^+e^-$&$1.19 \pm 0.20 $ &  $1.58 \pm 0.49$ \gh \\
\hline
$B\to K^*\mu^+\mu^-$&$1.06 \pm 0.09 $ & $1.19 \pm 0.39$ \gh \\ \hline
$B\to X_s \mu^+ \mu^-$ & $4.3 \pm 1.2$ &  $4.2\pm 0.7$ \gh
 \\ \hline
$B\to X_s e^+ e^-$ & $4.7 \pm 1.3$ &  $ 4.2\pm 0.7$ \gh \\ 
\hline
$B\to X_s \ell^+ \ell^-$ & $4.5 \pm 1.0 $ &  $ 4.2\pm 0.7$ \gh \\
\hline
\label{tab:exclusive-decays}
\end{tabular}}
\end{center}
\end{table}

\subsection{Angular analysis of the decay $B^0 \to K^{*0}( \to K^+ \pi^-) \mu^+ \mu^-$}

\begin{figure}
\centerline{\includegraphics[width=3.5in]{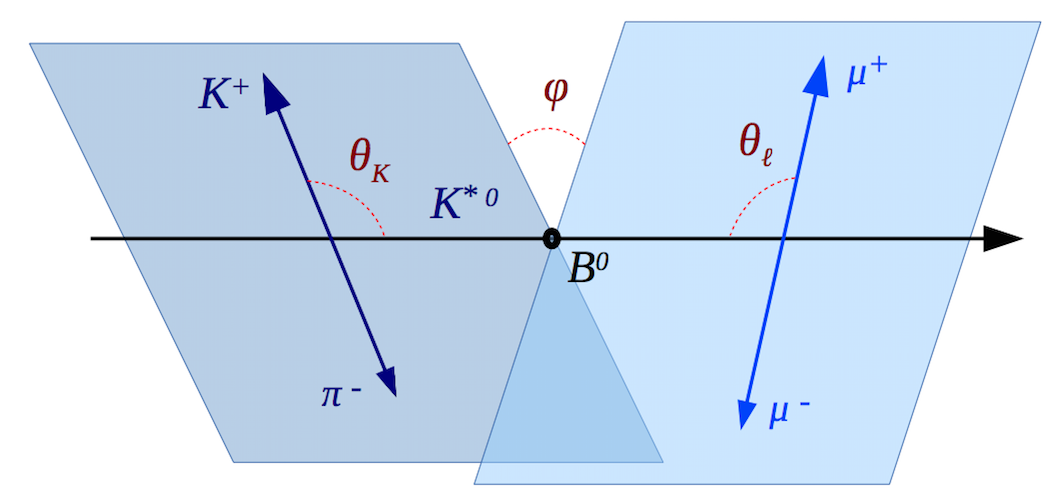}}
\caption{Definitions of the angles in $B^0 \to K^{*0}( \to K^+ \pi^-) \mu^+ \mu^-$.}
\label{fig:angular-analysis}
\end{figure}

For the inclusive decays $ B \to X_s \ell^+ \ell^-$, the observables which have been measured are the
integrated rates, the dilepton invariant mass  $d\Gamma/dq^2 $ and the FB asymmetry $A_{\rm FB}(q^2) $.
They are all found to be in agreement with the SM. In the exclusive decays such as $B \to K^* \ell^+ \ell^-$
and $B_s \to \phi \ell^+ \ell^-$, a complete  angular analysis of the decay is experimentally feasible.
This allows one to measure a number of additional observables, defined  below.  
\begin{eqnarray}
\frac{1}{d(\Gamma + \bar\Gamma)} \frac{d^4(\Gamma + \bar\Gamma)}{dq^2 d\Omega }
&=& \frac{9}{32 \pi}\left[\frac{3}{4} (1-F_L)\sin^2\theta_K + F_L \;\cos^2\theta_K \right. \nonumber \\
&& \left.+ \frac{1}{4} (1-F_L)\sin^2\theta_K\;\cos 2\theta_\ell - F_L \cos^2\theta_K \cos 2\theta_\ell \right. \nonumber \\
&& \left.+S_3 \sin^2 \theta_K\; \sin^2\theta_\ell\;\cos 2\phi + S_4 \sin 2\theta_K \sin 2 \theta_\ell \cos\phi\right. \nonumber \\
&& \left.+S_5 \sin 2\theta_K \sin \theta_\ell \cos\phi + \frac{4}{3}\; A_{\rm FB}\sin^2\theta_K \cos \theta_\ell\right. \nonumber \\
&& \left.+ S_7 \sin 2\theta_K \sin \theta_\ell\;\sin\phi + S_8 \sin 2\theta_K \sin 2\theta_\ell\;\sin\phi\right. \nonumber \\
&& \left.+ S_9 \sin^2 \theta_K  \sin^2\theta_\ell\;\sin 2\phi 
\right].
\label{eq:angular-observables}
\end{eqnarray}
The three angles   $\theta_K$, $ \theta_\ell $ and $ \phi $ for the decay $B^0 \to K^{*0}( \to K^+ \pi^-) \mu^+ \mu^-$
are  defined in Fig. \ref{fig:angular-analysis}.%
An angular analysis of  the decay chains
$B^0 \to K^{*0}( \to K^+ \pi^-) \mu^+ \mu^-$ \cite{Aaij:2015oid} and $B_s^0 \to  \phi ( \to K^+ K^-) \mu^+ \mu^-$
\cite{Aaij:2013aln}  has been carried out by LHCb.  

 The observables in (\ref{eq:angular-observables}) are
$q^2$-dependent coefficients of the Wilson coefficients and hence they probe the underlying dynamics.
Since these coefficients have been calculated to a high accuracy, the remaining theoretical
uncertainty lies in the form factors and also from the charm-quark loops. The form factors have been
calculated using the QCD sum rules and in the high-$q^2$ region also using lattice QCD. They limit the
current theoretical accuracy. However, a number of 
so-called optimised observables has been proposed \cite{Descotes-Genon:2013vna}, which reduce the dependence on the form factors. Using the
LHCb convention, these observables  are defined as \cite{Aaij:2015oid}
 \begin{eqnarray}
P_1 &\equiv& 2 S_3/(1- F_L);~~P_2 \equiv 2 A_{\rm FB}/ 3(1-F_L);~~P_3 \equiv -S_9/(1-F_L), \nonumber\\
&& P^\prime_{4,5,6,8}  \equiv  S_{4,5,7,8}/\sqrt{F_l(1-F_L)}.
\end{eqnarray}
\begin{figure}
\centerline{\includegraphics[width=4.5in]{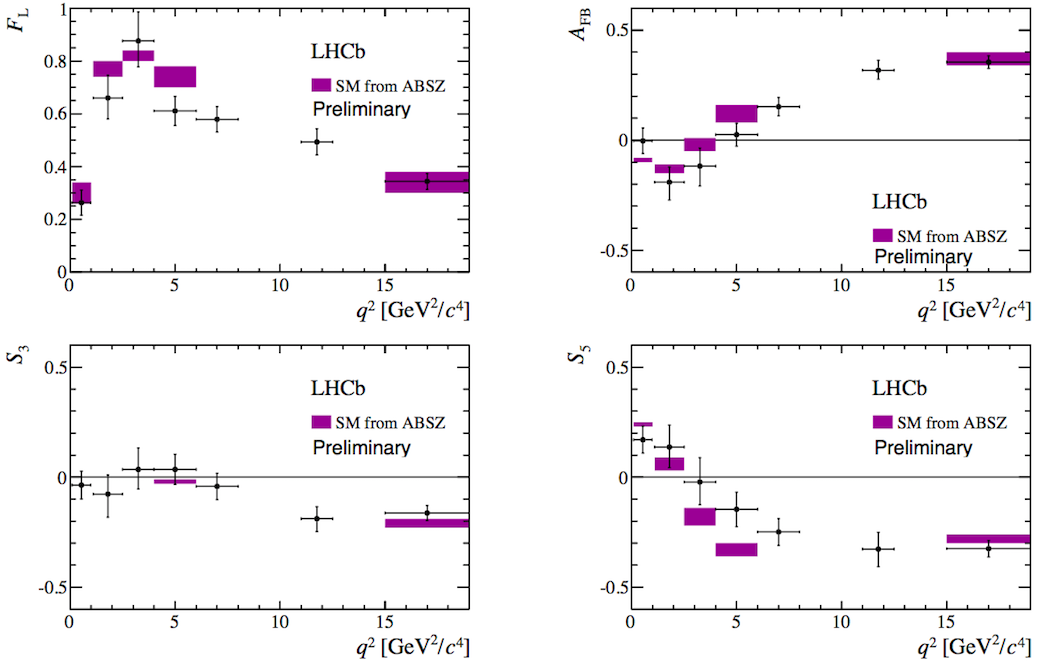}}
\caption{CP-averaged variables in bins of $q^2$ for the observables
 $F_{\rm L}$, $A_{\rm FB}$, 
$S_3$ and $S_5$ in $B^0 \to K^{*0}( \to K^+ \pi^-) \mu^+ \mu^-$ measured by LHCb \cite{Aaij:2015oid}  and
comparison with the SM  \cite{Altmannshofer:2014rta}.} 
\label{fig:lhcb-ang-analys-results-1}
\end{figure}
\begin{figure}
\centerline{\includegraphics[width=4.5in]{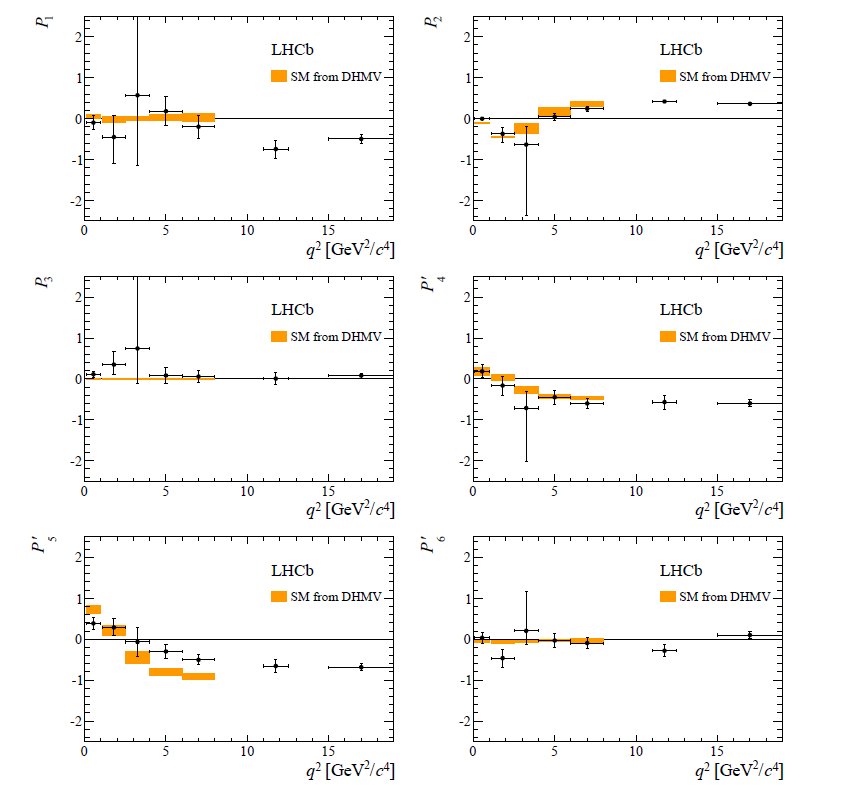}}
\caption{The optimised angular observables in  bins of $q^2$  in the decay $B^0 \to K^{*0}( \to K^+ \pi^-)\; \mu^+ \mu^-$
measured by the LHCb collaboration  \cite{Aaij:2015oid} and comparison with the SM  \cite{Descotes-Genon:2014uoa}.}
\label{fig:lhcb-ang-analys-results-2}
\end{figure}
These angular observables have been analysed in a number of theoretical studies
\cite{Descotes-Genon:2014uoa,Jager:2012uw,Jager:2014rwa,Altmannshofer:2014rta,Straub:2015ica,Hurth:2013ssa},
which differ in the treatment of their non-perturbative input, mainly form factors.                                   
 The LHCb collaboration, which currently
dominates this field, has used these SM-based estimates and compared with their data in various
$q^2$ bins. Two representative comparisons based on the theoretical estimates from
 Altmannshofer and Straub \cite{Altmannshofer:2014rta} and Descotes-Genon, Hofer, Matias and 
 Virto  \cite{Descotes-Genon:2014uoa} are shown in Figs. \ref{fig:lhcb-ang-analys-results-1} and
 \ref{fig:lhcb-ang-analys-results-2}, respectively. They are largely in agreement with the data, except for the
 distributions in the observables $S_5(q^2)$  (in Fig. \ref{fig:lhcb-ang-analys-results-1}) and
  $P_5^\prime(q^2)$ (in Fig. \ref{fig:lhcb-ang-analys-results-2})
   in the bins around $q^2 \geq 5$ GeV$^2$.
 The pull on the SM depends on the theoretical model, reaching 3.4$\sigma$ in the bin
 $4.3 \leq q^2 \leq 8.68$ GeV$^2$ compared to DHMV \cite{Descotes-Genon:2014uoa}. 
  There are deviations of a similar nature, between 2 and 3$\sigma$,  seen in
 the comparison of  $S_5$ and other quantities, such as the partial branching ratios in $B \to K^* \mu^+ \mu^-$,
 $B _s \to  \phi  \mu^+ \mu^-$ and $F_L(q^2)$ \cite{Altmannshofer:2014rta}.

\begin{figure}
\centerline{\includegraphics[width=5.5in]{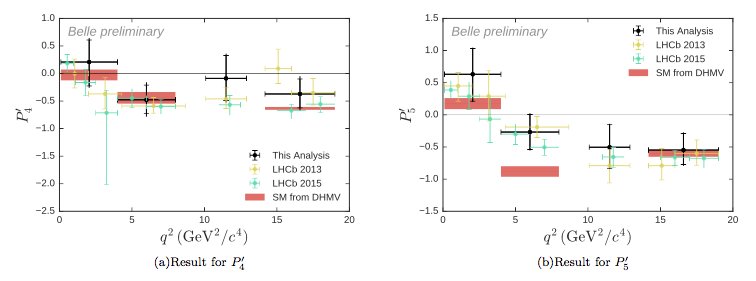}}
\caption{The optimised angular observables $P_4^\prime $ and $P_5^\prime $
in  bins of $q^2$  in the decay $B^0 \to K^{*0}( \to K^+ \pi^-)\; \mu^+ \mu^-$
measured by  Belle \cite{Abdesselam:2016llu}  and comparison with the SM  \cite{Descotes-Genon:2014uoa}.}
\label{fig:belle-ang-analys-results-2}
\end{figure}

     An analysis of the current Belle data \cite{Abdesselam:2016llu} , shown in Fig. \ref{fig:belle-ang-analys-results-2}, 
 displays a similar pattern as the one reported by LHCb. As the Belle data has larger errors,
 due to limited statistics, the resulting pull on the SM is less significant. In the interval 
 $4.0 \leq q^2 \leq 8.0$ GeV$^2$, Belle reports deviations of $2.3\sigma$ (compared to DHMV \cite{Descotes-Genon:2014uoa}), $1.72\sigma$ (compared to BSZ \cite{Straub:2015ica}) and $1.68 \sigma$ (compared to
 JC \cite{Jager:2014rwa}). These measurements will improve greatly at Belle II.
 
\begin{figure}
\centerline{\includegraphics[width=3.5in]{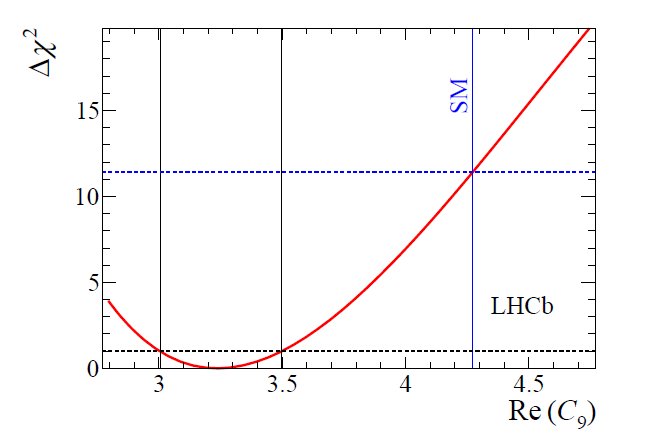}}
\caption{The $\Delta \chi^2$ distribution for the real part of the Wilson coefficient $ {\rm Re} C_9(m_b) $
from a fit of the CP-averaged observables $F_{\rm L}, A_{\rm FB}, S_3, ...,S_9$ in $B^0 \to K^{*0}( \to K^+ \pi^-) \mu^+ \mu^-$ by the LHCb collaboration \cite{Aaij:2015oid}.}
\label{fig:lhcb-chi-square}
\end{figure}

  To quantify the deviation of the LHCb data from the SM estimates,
 a $\Delta \chi^2$ distribution for the real part of the Wilson coefficient $ {\rm Re} C_9(m_b) $ is shown in Fig.
 \ref{fig:lhcb-chi-square}. In  calculating the  $\Delta \chi^2$, the other Wilson coefficients are set to their SM values. 
 The coefficient $ {\rm Re} C_9^{\rm SM}(m_b)=4.27 $ at the NNLO accuracy  in the SM is indicated by a vertical line. 
  The best fit of the LHCb data yields a value which is shifted from the SM, and the deviation in this coefficient
 is found to be $ \Delta {\rm Re} C_9 (m_b)=-1.04 \pm 0.25 $. The deviation is tantalising, but not yet
 conclusive. A bit of caution is needed here as the SM estimates used in the analysis above may have to be revised, once the residual
 uncertainties are better constrained. In particular, the hadronic contributions generated by the four-quark operators with charm are difficult to estimate, especially around $q^2 \sim 4 m_c^2$, leading to an effective shift in the 
 value of the Wilson coefficient being discussed \cite{Ciuchini:2015qxb}.
 
\section{CKM-suppressed $ b \to d \ell^+ \ell^- $ transitions in the SM}

Weak transitions $ b \to d \ell^+ \ell^- $ , like the radiative decays $b \to d \gamma$, are CKM suppressed and because of this the structure of the
effective weak Hamiltonian is different than the one encountered earlier for the $ b \to s \ell^+ \ell^- $
transitions.

\begin{eqnarray}
{\cal H}_{\rm eff}^{b \to d} &=&  - \frac{4 G_G }{\sqrt{2}}\left[V_{tb}^* V_{td} \sum_{i=1}^{10}
C_i(\mu)(O_i(\mu)\right. \nonumber \\
 && \left. + V_{ub}^* V_{ud} \sum_{i=1}^{2} C_i(\mu)( O_i(\mu) - O_i(\mu) )\right] + {\rm h.c.}.
\end{eqnarray}
Here $O_i(\mu)$ are the dimension-six operators introduced earlier (except for the interchange $s \to d$
quark) in $ {\cal H}_{\rm eff}^{b \to s}$. As the two CKM factors 
are comparable in magnitude $\vert V_{tb}^* V_{td}\vert  \simeq \vert V_{ub}^* V_{ud} \vert $, and have different weak phases,
we  anticipate sizeable CP-violating asymmetries in both the inclusive $ b \to d \ell^+ \ell^- $ and
exclusive transitions, such as $B \to (\pi, \rho) \ell^+ \ell^-$. 
The relevant operators appearing in $ {\cal H}_{\rm eff}^{b \to d}$ are:

\underline{Tree operators} 
\begin{equation}
{\cal O}_1 =  
\left ( \bar d_L \gamma_\mu T^A c_L \right ) 
\left ( \bar c_L \gamma^\mu T^A b_L \right ), \quad
{\cal O}_2 = 
\left ( \bar d_L \gamma_\mu c_L \right ) 
\left ( \bar c_L \gamma^\mu b_L \right ) , 
\end{equation}

\begin{equation}
{\cal O}_1^{(u)} =  
\left ( \bar d_L \gamma_\mu T^A u_L \right ) 
\left ( \bar u_L \gamma^\mu T^A b_L \right ), \quad
{\cal O}_2^{(u)} = 
\left ( \bar d_L \gamma_\mu u_L \right ) 
\left ( \bar u_L \gamma^\mu b_L \right ). 
 \end{equation}
\underline{Dipole operators}
\begin{equation}
{\cal O}_7 = 
\frac{e \, m_b}{g_{\rm s}^2}   
\left ( \bar d_L \sigma^{\mu \nu} b_R \right ) F_{\mu \nu}, \quad 
{\cal O}_8 = 
\frac{m_b}{g_{\rm s}}   
\left ( \bar d_L \sigma^{\mu \nu} T^A b_R \right ) G_{\mu \nu}^A. 
\end{equation}
\underline{Semileptonic operators}
\begin{equation}
{\cal O}_9 = 
\frac{e^2}{g_{\rm s}^2}   
\left ( \bar d_L \gamma^\mu b_L \right ) 
\sum_\ell \left ( \bar \ell \gamma_\mu \ell \right ), \quad
{\cal O}_{10} = 
\frac{e^2}{g_{\rm s}^2}   
\left ( \bar d_L \gamma^\mu b_L \right ) 
\sum_\ell \left ( \bar \ell \gamma_\mu \gamma_5 \ell \right ).
\end{equation}
Here, $e(g_s)$ is the QED (QCD) coupling constant. Since the inclusive decay $B \to X_d \ell^+ \ell^-$ has
not yet been measured, but hopefully will be at Belle II, we discuss the exclusive decay $B^+ \to \pi^+ \ell^+ \ell^- $, which
is the only $b \to d$ semileptonic transition measured so far.

\subsection{Exclusive decay $B^+ \to \pi^+  \ell^+ \ell^- $}

The  decay \footnote{Charge conjugation is implied here.} $B^+ \to \pi^+ \ell^+ \ell^- $  is induced
by the vector and tensor currents and their matrix elements are defined as  
\begin{equation}
\langle \pi (p_\pi) | \bar b \gamma^\mu d | B (p_B) \rangle =
 f^\pi_+ (q^2) \left ( p_B^\mu + p_\pi^\mu \right ) +  
\left [  f^\pi_0 (q^2) -  f^\pi_+ (q^2) \right ]   
\displaystyle\frac{m^2_B - m^2_\pi}{q^2} q^\mu,
\end{equation}
 %
 \vspace*{-6mm}
 \begin{equation}
\langle  \pi (p_\pi) | \bar b \sigma^{\mu \nu} q_\nu d | B (p_B) \rangle =
\displaystyle\frac{ i  f^\pi_T (q^2)} {m_B + m_\pi} 
\left [ \left ( p_B^\mu + p_\pi^\mu \right ) q^2 - 
q^\mu \left ( m_B^2 - m_\pi^2 \right ) \right ]. 
\end{equation}
These form factors are related to the ones in the decay $B \to K \ell^+ \ell^-$, called $f^K_i(q^2) $,
 discussed earlier, by $SU(3)_F$
symmetry. Of these, the form factors $f^\pi_+(q^2)  $ and $f^\pi_0(q^2) $ are related by isospin symmetry to the corresponding ones
measured in the charged current process $B^0 \to \pi^- \ell^+ \nu_\ell $ by Babar and Belle, and they can
be extracted from the data. This has been done  using several parameterisations of the form factors with all of them giving an adequate description of the data \cite{Ali:2013zfa}. Due to their analytic properties, the so-called
$z$-expansion methods, in which the form factors are expanded in a Taylor series in $z$, employed in the
Boyd-Grinstein-Lebed (BGL) parametrisation \cite{Boyd:1994tt} and the Bourrely-Caprini-Lellouch (BCL) \cite{Bourrely:2008za} parametrisation, are preferable.

The BGL parametrisation is used in working out the  decay rate and  the invariant dilepton mass
 distribution \cite{Ali:2013zfa} for  $B^+ \to \pi^+ \ell^+ \ell^- $, which is discussed below.
  The BCL-parametrisation is used by
 the lattice-QCD groups, the HPQCD \cite{Bouchard:2013zda,Bouchard:2013pna}
  and Fermilab/MILC \cite{Bailey:2015nbd} collaborations, to determine the form factors
 $f^\pi_i(q^2) $ and $f^K_i(q^2)$.  In particular, the Fermilab/MILC collaboration has worked out the dilepton
 invariant mass distribution in the decay of interest $B^+ \to \pi^+ \ell^+ \ell^- $, making use of their
 simulation in the large-$q^2$ region and extrapolating with the BCL parametrisation. 
 
 We first discuss the low-$q^2$ region ($q^2 \ll m_b^2$).  In this case,, heavy quark symmetry (HQS) relates all three
form factors of interest $f^\pi_i(q^2) $
 and this can be used advantageously to have a reliable estimate of the dilepton invariant mass
spectrum in this region. Including lowest order HQS-breaking, the resulting expressions
for the form factors  (for $q^2/m_b^2 \ll 1 $) are
worked out  by Beneke and Feldmann \cite{Beneke:2000wa}.
Thus,  fitting  the form factor $f_+ (q^2)$ from the charged current data on 
 $B \to \pi \ell^+ \nu_\ell$ decay, and taking into account the HQS and its breaking, lead to a  model-independent predictions of the differential  branching ratio (dimuon mass spectrum)  in  the neutral current process $ B^+ \to \pi^+ \ell^+ \ell^-$ for low-$q^2$ values. However, the long-distance contribution, which arises from the
 processes $B^+ \to \pi^+ (\rho^0, \omega) \to \pi^+ \mu^+ \mu^-$ are not included here.  
The SM invariant dilepton mass distribution in $ B^+ \to \pi^+ \ell^+ \ell^-$ 
integrated over the range $1 {\rm GeV}^2 \leq q^2 \leq  8 {\rm GeV}^2$
 yields a partial branching ratio  
\begin{equation}
{\cal B}(B^+ \to \pi^+ \mu^+ \mu^-) = (0.57 ^{+0.07}_{-0.05}) \times 10^{-8}. 
\end{equation}
Thanks to the available data on the charged current process and heavy quark symmetry, this enables an
accuracy of about 10\% for an exclusive branching ratio, comparable to the theoretical accuracy in the
inclusive decay $B \to X_s \gamma$, discussed earlier. Thus,
the decay $ B^+ \to \pi^+ \mu^+ \mu^- $ offers 
 a key advantage compared to the decay $ B^+ \to K^+ \ell^+ \ell^-$, in which case the
charged current process is not available. 

 The differential branching ratio in the entire $q^2$ region is given by
 \begin{equation}
\frac{ d {\cal B}(B^+ \to \pi^+ \ell^+ \ell^-)}{d q^2} = C_B \vert V_{tb}V_{td}^* \vert^2 \sqrt{\lambda(q^2)}
\sqrt{1- \frac{4 m_\ell^2}{q^2}} F(q^2),
\end{equation}
 where the constant $C_B=G_F^2 \alpha_{\rm em}^2 \tau_B/1024 \pi^5 m_B^3$ and $ \lambda(q^2)$ is the
 usual kinematic function $\lambda(q^2)= (m_B^2 + m_\pi^2 -q^2)^2 - 4 m_B^2 m_\pi^2 $. The function
 $F(q^2)$ depends on the effective Wilson coefficients, $C_7^{\rm eff}$, $C_9^{\rm eff}$, and $C_{10}^{\rm eff}$,
 and the three form factors $f_+^\pi (q^2)$, $f_0^\pi (q^2)$ and $f_T^\pi (q^2)$.
 A detailed discussion of the determination of the form factors, of which only $f_+^\pi (q^2)$ and $f_T^\pi (q^2)$
 are numerically important for $\ell^\pm = e^\pm, \mu^\pm$  is given elsewhere \cite{Ali:2013zfa}. We recall that
 $f_+^\pi (q^2)$ is constrained by the data on the charged current process in the entire
 $q^2$ domain. In addition, the lattice-QCD results on the form factors  in the large-$q^2$ domain
 and the HQS-based relations in the low-$q^2$ region provide sufficient constraints on the form factor.
 This has enabled a rather precise determination of the invariant dilepton mass distribution 
 in $B^+ \to \pi^+ \ell^+ \ell^-$.
 
\begin{figure}
\centerline{\includegraphics[width=4.5in]{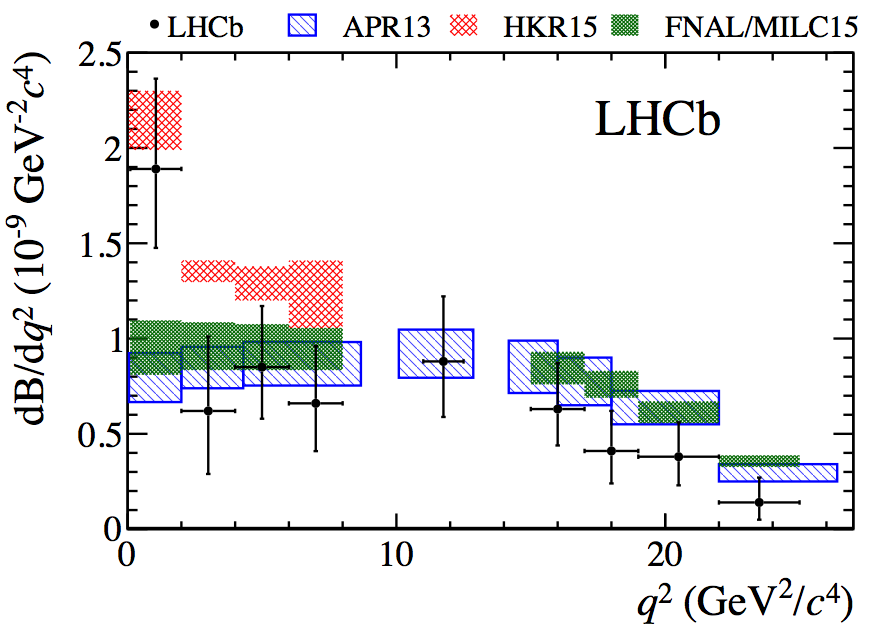}}
\caption{Comparison of  the dimuon invariant mass distribution
 in $B^+ \to \pi^+  \mu^+ \mu^-$ in the SM with the LHCb data \cite{Aaij:2015nea}.
 Theoretical distributions shown are : APR13 \cite{Ali:2013zfa}, HKR15 \cite{Hambrock:2015wka},
 and FNAL/MILC \cite{Bailey:2015nbd}. }
\label{fig:lhcb-Bpimumu-result}
\end{figure}

 Taking into account the various parametric and form-factor dependent uncertainties, this yields the following
 estimate for the branching ratio for $B^+ \to \pi^+ \mu^+ \mu^- $ \cite{Aaij:2015nea}
\begin{equation}
{\cal B}_{\rm SM}(B^+ \to \pi^+ \mu^+ \mu^-) = (1.88 ^{+0.32}_{-0.21}) \times 10^{-8}, 
\label{eq:Bpimu}
\end{equation}
to be compared with the measured branching ratio by the LHCb collaboration \cite{Aaij:2015nea}
 (based on 3${\rm fb}^{-1}$ data):
\begin{equation}
{\cal B}_{\rm LHCb}(B^+ \to \pi^+ \mu^+ \mu^-) = (1.83 \pm 0.24 \pm 0.05) \times 10^{-8}, 
\label{eq:BpimuLHCb}
\end{equation}
where the first error is statistical and the second systematic, resulting in excellent agreement.
The dimuon invariant mass distribution measured by the LHCb collaboration \cite{Aaij:2015nea}
 is shown in Fig. \ref{fig:lhcb-Bpimumu-result}, and compared with 
the SM-based theoretical prediction, called APR13 \cite{Ali:2013zfa}, and the
lattice-based  calculation, called FNAL/MILC 15 \cite{Bailey:2015nbd}.
Also shown is a comparison with a calculation, called HKR \cite{Hambrock:2015wka} , which has 
essentially the same short-distance contribution in the low-$q^2$ region, as discussed earlier, but additionally takes into account the contributions from the lower resonances $\rho, \omega$ and $\phi$.
This adequately describes the distribution in the $q^2$ bin, around
1 GeV$^2$.

 With the steadily improving lattice calculations for the various input hadronic quantities and the
form factors, theoretical error indicated in Eq. (\ref{eq:Bpimu}) will go down considerably. Experimentally,
we expect rapid progress due to the increased statistics at the LHC, but also from  Belle II, which will 
measure the corresponding distributions and branching ratio also in the decays $ B^+ \to \pi^+ e^+ e^- $,
and  $B^+ \to \pi^+ \tau^+ \tau^-  $, providing a complementary test of
the $e$-$\mu$-$\tau$ universality in $b \to d$ semileptonic transitions. 
 
\section{Leptonic Rare $B$ Decays}

The final topic discussed in this write-up involves purely leptonic decays $ B_s^0  \to \ell^+ \ell^-$ 
and  $ B^0  \to \ell^+ \ell^-$  with $\ell^+ \ell^- =e^+ e^-, \mu^+ \mu^-, \tau^+ \tau^- $. Of these, 
$ {\cal B}(B_s^0 \to \mu^+ \mu^-)= (2.8^{+0.7}_{-0.6}) \times 10^{-9} $ is now well measured, and the corresponding CKM-suppressed decay  $ {\cal B}(B^0 \to \mu^+ \mu^-)= (3.9^{+1,6}_{-1.4}) \times 10^{-10} $ is almost on the verge
of becoming a measurement. These numbers are from the combined CMS/LHCb
data \cite{CMS:2014xfa}. From the experimental point of view, their measurement is
a real {\it tour de force}, considering the tiny branching ratios and the formidable background at the LHC. 

In the SM, these decays are dominated by the axial-vector operator
 $O_{10} = (\bar{s}_\alpha \gamma^\mu P_L b_\alpha) (\bar{\ell} \gamma_\mu \gamma_5 \ell) $.
 In principle,  the operators $O_S= m_b (\bar{s}_\alpha \gamma^\mu P_R b_\alpha) (\bar{\ell} \ell)$ and 
 $O_P= m_b (\bar{s}_\alpha \gamma^\mu P_R b_\alpha) (\bar{\ell} \gamma_5 \ell)$ also contribute, but  are
 chirally suppressed in the SM. This need not be the case in BSM scenarios, and hence the
 great interest in measuring these decays precisely. In the SM, the measurement of $ {\cal B}(B_s^0 \to \mu^+ \mu^-)$ and  $ {\cal B}(B^0 \to \mu^+ \mu^-)$  provide a measurement of the
 Wilson coefficient $C_{10} (m_b)$. Their ratio
 $ {\cal B}(B^0 \to \mu^+ \mu^-)/{\cal B}(B_s^0 \to \mu^+ \mu^-) $ being proportional to the ratio of the
 CKM matrix-elements $\vert V_{td}/V_{ts}\vert^2$ is an important constraint on the CKM unitarity triangle.
 \begin{figure}
\centerline{\includegraphics[width=4.5in]{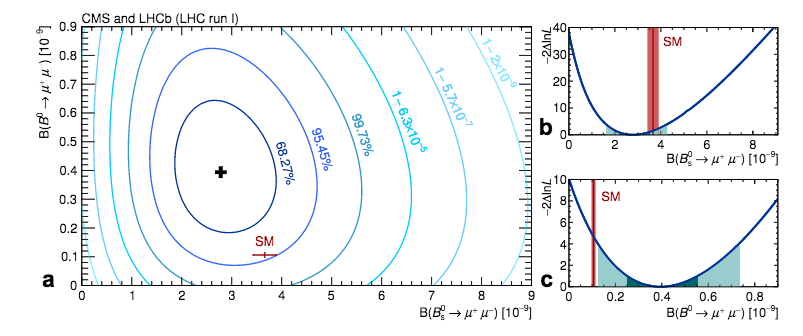}}
\caption{Likelihood contours in the ${\cal B}(B^0 \to \mu^+ \mu^-)$ versus ${\cal B}(B_s^0 \to \mu^+ \mu^-)$  
plane. The (black) cross in (a) marks the best-fit value and the SM expectation is shown as the (red) marker.
Variations of the test statistics $-2 \Delta \ln L$ for ${\cal B}(B_s^0 \to \mu^+ \mu^-)$ (b) and
${\cal B}(B^0 \to \mu^+ \mu^-)$ (c) are shown. The SM prediction is denoted with the vertical (red) bars.
(From the combined CMS-LHCb data \cite{CMS:2014xfa}.) }
\label{fig:LHCb-CMS-mumu}
\end{figure}

 The decay rate $\Gamma(B_s^0 \to \mu^+ \mu^-)$  in the SM can be written as 
\begin{equation}
\Gamma(B_s^0 \to \mu^+ \mu^-)= \frac{G_F^2 M_W^2 m_{B_s}^3 f_{B_s}^2} { 8 \pi^5} \vert V_{tb}^*V_{ts}\vert^2
\frac{4 m_\ell^2}{ m_{B_s}^2}  \sqrt{1- \frac{4 m_\ell^2}{ m_{B_s}^2} } \vert C_{10}\vert^2 + O(\alpha_{\rm em}).
\end{equation}
The coefficient $C_{10} $ has been calculated  by taking into account the NNLO QCD corrections and
NLO electroweak corrections, but the $O(\alpha_{\rm em})$ contribution indicated above is ignored, as it is small. The SM branching
ratio in this accuracy have been obtained in \cite{Bobeth:2013uxa,Hermann:2013kca,Bobeth:2013tba}, where a careful account of the various
input quantities is presented. The importance of including the effects of the width difference $\Delta \Gamma_s$ 
due to the $B_s^0$ - $\bar{B}_s^0$ mixings in extracting the branching ratio for $B_s \to \mu^+ \mu^-$
has been emphasised in the literature \cite{DeBruyn:2012wk} and is included in the analysis.
The time-averaged branching ratios, which in the SM to a good approximation equals to
 $\overline {\cal B} (B_s \to \mu^+ \mu^-)= \Gamma( B_s \to \mu^+ \mu^-)/\Gamma_H (B_s)$, where  
$\Gamma_H (B_s)  $ is the heavier mass-eigenstate total width, is given below \cite{Bobeth:2013uxa}
\begin{equation}
\overline {\cal B} (B_s \to \mu^+ \mu^-) =(3.65 \pm 0.23) \times 10^{-9}.
\label{eq:Bobeth-Bsmumu}
\end{equation}
In  evaluating this, a value $f_{B_s}=227.7(4.5)$ MeV  was used   from the earlier FLAG average \cite{Aoki:2013ldr}. In the most recent
compilation by the FLAG collaboration \cite{Aoki:2016frl} , this coupling constant has been updated to
$f_{B_s}=224(5)$ MeV, which reduces the branching ratio to $\overline {\cal B} (B_s \to \mu^+ \mu^-) =(3.55 \pm 0.23) \times 10^{-9}$. This is compatible with the current measurements to about 1$\sigma$, with the uncertainty
dominated by the experiment.

 The corresponding branching ratio  $\overline {\cal B} (B^0 \to \mu^+ \mu^-)$ is
evaluated as  \cite{Bobeth:2013uxa}
\begin{equation}
\overline {\cal B} (B^0 \to \mu^+ \mu^-) =(1.06 \pm 0.09) \times 10^{-10},
\label{eq:Bobeth-Bdmumu}
\end{equation}
which, likewise, has to be scaled down to  $ (1.01 \pm 0.09) \times 10^{-10} $, due to the current 
average \cite{Aoki:2016frl}
$f_{B}=186(4)$ MeV, compared to $f_{B}=190.5(4.2)$ MeV used in deriving the result given in 
Eq. (\ref{eq:Bobeth-Bdmumu}). This is about 2$\sigma$ below the current measurement, and the ratio of the
two leptonic decays $\overline {\cal B} (B_s \to \mu^+ \mu^-)/\overline {\cal B} (B^0 \to \mu^+ \mu^-) $ is
off by about 2.3$\sigma$. The likelihood contours in the ${\cal B}(B^0 \to \mu^+ \mu^-)$ versus ${\cal B}(B_s^0 \to \mu^+ \mu^-)$  plane from the combined CMS/LHCb data are shown in Fig. \ref{fig:LHCb-CMS-mumu}. 

The anomalies  in the decays $ B \to K^* \mu^+ \mu^-$, discussed previously, and the deviations in 
$ {\cal B}(B^0 \to \mu^+ \mu^-)$  and ${\cal B}(B_s^0 \to \mu^+ \mu^-)$, if consolidated experimentally,
would require an extension of the SM. A recent proposal based on the group $SU(3)_C \times SU(3)_L \times U(1)$
is discussed by Buras, De Fazio and Girrbach \cite{Buras:2013dea}. Lepton non-universality, if confirmed,
requires a leptoquark-type solution. A viable candidate theory to replace the SM and accounting for all the
current anomalies, in my opinion, is not in sight.

\section{Global fits of the Wilson Coefficients $C_9$ and $C_{10}$}

As discussed in the foregoing, a number of deviations from the SM-estimates are currently present in the
data on semileptonic and leptonic  rare $B$-decays. They lie mostly around 2 to 3$\sigma$.
 A comparison of the LHCb data on a number of angular observables 
$F_{\rm L}, A_{\rm FB}, S_3, ...,S_9$ in $B^0 \to K^{*0}( \to K^+ \pi^-) \mu^+ \mu^-$ with the SM-based
estimates was shown in Fig, \ref{fig:lhcb-chi-square}, yielding a value of ${\rm Re}( C_9)$ which deviates
from the SM by about 3$\sigma$. A number of groups has undertaken similar fits of the data and the outcome
depends on a number of assumed correlations. However, it should be stressed that there are still
non-perturbative  contributions present in the current theoretical estimates which are not
yet under complete quantitative control. The contributions from the charm quarks in the loops is a case in
point. Also,  form factor uncertainties are probably larger than assumed in some of
these global fits. 

\begin{figure}
\centerline{\includegraphics[width=4.5in]{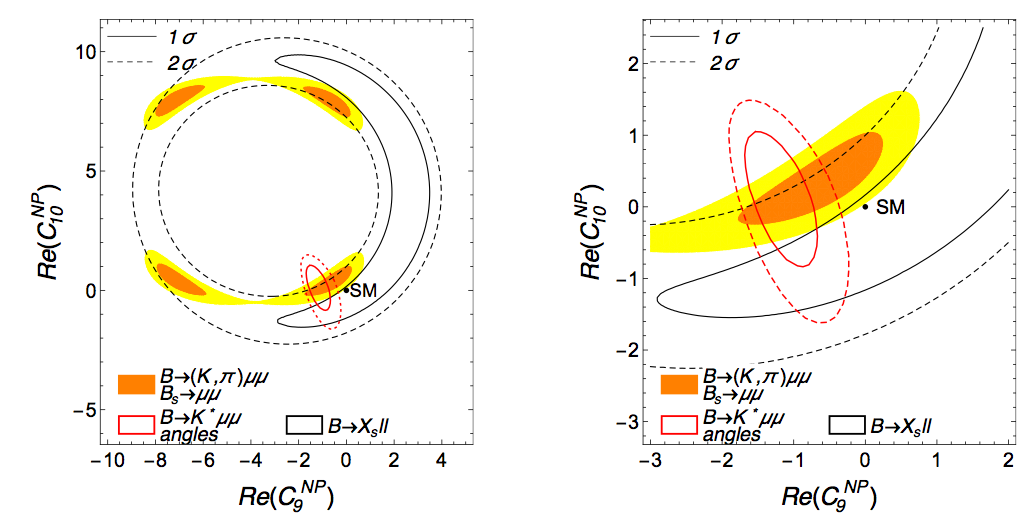}}
\caption{Present constraints on the Wilson coefficients ${\rm Re} C_{10}^{\rm NP}$ vs. 
${\rm Re} C_{9}^{\rm NP}$ from the semileptonic rare $B$-decays and $B_s \to \mu^+ \mu^-$.
The SM-point is indicated.
(From Fermilab/MILC Lattice Collaboration \cite{Bailey:2015nbd}.)}
\label{fig:fermilab-milc-c9-10}
\end{figure}

As a representative example of the kind of constraints  on the Wilson coefficients $C_9$ and $C_{10}$ that
follow from the data on semileptonic and leptonic decays of the $B$ mesons is shown in
 Fig. \ref{fig:fermilab-milc-c9-10} from the Fermilab/MILC collaboration \cite{Bailey:2015nbd}. This shows that the SM point indicated by
 $(0,0)$ in the ${\rm Re}(C_{9}^{NP},{\rm Re}(C_{10}^{NP})$-plane lies a little beyond 2$\sigma$. In some other
 fits, the deviations are larger but still far short for a discovery of BSM effects.
 As a lot of the experimental input in this and similar analysis is due to the LHCb data, this has to be confirmed
 by an independent experiment. This, hopefully, will be done by Belle II. We are better advised to wait and see
 if these deviations become statistically significant enough to warrant new physics. Currently, the situation is
 tantalising but not conclusive.

\section{Concluding Remarks}

From the measurement by the CLEO collaboration of the rare decay $B \to X_s \gamma$ in 1995,
having a branching ratio of about $3 \times 10^{-4}$, to the rarest of  the measured $B$ decays, $B^0 \to \mu^+ \mu^-$,
with a branching fraction of about $1 \times 10^{-10}$ by the LHCb and CMS collaborations,  SM has been
tested over six orders of magnitude. This is an impressive feat,  made possible by dedicated experimental
programmes carried out with diverse beams and detection techniques over a period of more than 20 years.
A sustained theoretical effort has accompanied the experiments all along, underscoring both the continued
theoretical interest in $b$ physics and an intense exchange between the two communities.  With the exception of a
few anomalies, showing deviations from the SM ranging between 2 to 4$\sigma$ in statistical significance,
a vast majority of  the measurements is in quantitative agreement with the SM. In particular, all  quark flavour
transitions are described by the CKM matrix whose elements are now determined. The CP asymmetry
measured so far in laboratory experiments is explained by the Kobayashi-Maskawa phase. FCNC processes,
of which rare $B$-decays discussed here is a class, are governed by the GIM mechanism, with the particles
in the SM (three families of quarks and leptons, electroweak gauge bosons, gluons, and the Higgs) accounting for all the observed phenomena - so far. Whether this astounding consistence will continue will be tested in the
 coming years, as the LHC experiments  analyse more data, enabling vastly improved precision in some
of the key measurements discussed here. In a couple of years from now, Belle II will start taking data, providing
independent and new measurements. They will be decisive in either deepening our knowledge about the SM,
or hopefully in discovering the new frontier of physics.

\section{Acknowledgment}

I thank Harald Fritzsch  for inviting me to this very stimulating conference and Prof. K.K. Phua for the
warm hospitality in Singapore. I also thank Mikolaj Misiak for reading the manuscript and helpful suggestions.

\end{document}